\documentclass{elsart}

\usepackage{amssymb}
\usepackage{amsmath}
\usepackage{color,soul}
\usepackage{bbm}
\usepackage{graphicx}     
\usepackage{nicematrix}
\usepackage{natbib}
\usepackage{adjustbox}
\usepackage{multirow}
\usepackage{array}
\usepackage{booktabs}
\usepackage{mathtools}
\usepackage{multicol}

\makeatletter
\def\ps@copyright{%
 \let\@oddhead\@empty
 \let\@evenhead\@empty
 \def\@oddfoot{\textit{Submitted Preprint} \hfill}%
 \let\@evenfoot\@oddfoot}
\makeatother

\begin{document}

\begin{frontmatter}
\title{YANNs: Y-wise Affine Neural Networks for Exact and Efficient Representations of Piecewise Linear Functions}

\thanks[footnoteinfo]{$^*$Corresponding author}

\author[WVU]{Austin Braniff}\ead{austin.braniff@mail.wvu.edu},
\author[WVU]{Yuhe Tian}$^{,*}$\ead{yuhe.tian@mail.wvu.edu}

\address[WVU]{Department of Chemical and Biomedical Engineering, West Virginia University, United States of America}  


\begin{abstract}
This work formally introduces Y-wise Affine Neural Networks (YANNs), a fully-explainable network architecture that continuously and efficiently represent piecewise affine functions with polytopic subdomains. Following from the proofs, it is shown that the development of YANNs requires no training to achieve the functionally equivalent representation. YANNs thus maintain all mathematical properties of the original formulations. Multi-parametric model predictive control is utilized as an application showcase of YANNs, which theoretically computes optimal control laws as a piecewise affine function of states, outputs, setpoints, and disturbances. With the exact representation of multi-parametric control laws, YANNs retain essential control-theoretic guarantees such as recursive feasibility and stability. This sets YANNs apart from the existing works which apply neural networks for approximating optimal control laws instead of exactly representing them. By optimizing the inference speed of the networks, YANNs can evaluate substantially faster in real-time compared to traditional piecewise affine function calculations. Numerical case studies are presented to demonstrate the algorithmic scalability with respect to the input/output dimensions and the number of subdomains. YANNs represent a significant advancement in control as the first neural network-based controller that inherently ensures both feasibility and stability. Future applications can leverage them as an efficient and interpretable starting point for data-driven modeling/control.
\end{abstract}

\end{frontmatter}

\section{Introduction}
Recent advancements in Machine Learning (ML) have had a significant impact on the field of control, notably through the use of Neural Networks (NNs) to efficiently approximate nonlinear functions \citep{hornikMultilayerFeedforwardNetworks1989,daoutidisMachineLearningProcess2024a,braniffRealtimeProcessSafety2025}. State-of-the art ML-based control approaches include data-driven control policies \citep{drgonaDifferentiablePredictiveControl2022,wangTutorialReviewPolicy2025}, model predictive control (MPC) with surrogate modeling \citep{neufangSurrogateBasedOptimizationTechniques2024}, etc. These provide instrumental methods particularly for the cases when a mechanistic model is not available \citep{yooReinforcementLearningBatch2021}, too complex to use in real-time \citep{braniffHierarchicalMultiparametricProgramming2024b,katzIntegrationExplicitMPC2020a}, or with fast system dynamics \citep{xuConstraintsInformedNeuralLaguerreApproximation2024,stogiannosModelPredictiveControl2018}.

However, an outstanding challenge of ML-based control is lacking the desired mathematical guarantees that are central to a control-theoretic approach \cite{turanClosedloopOptimisationNeural2024,fabianiReliablyStabilizingPiecewiseAffineNeural2023}. For example, it is difficult to guarantee stability when using ML techniques \citep{nakamura-zimmererNeuralNetworkOptimal2022}. This stems from the common issue that NN models are largely uninterpretable \citep{fanInterpretabilityArtificialNeural2021}, which has rendered them the term ``black box models" \citep{hassijaInterpretingBlackBoxModels2024}. The absence of rigorous mathematical assurance makes it difficult to reliably implement these techniques into control, especially for safety-critical systems \citep{drgonaSafePhysicsInformedMachine2025}. Several approaches have been developed to recover these guarantees or bring some form of trust in the NN-based control policy. These include: input to state stability (ISS) for linear systems \citep{kumarIndustrialLargescaleModel2021f}, using robust control methods to certify stability for networks with specific activation functions \citep{nguyenStabilityCertificatesNeural2021} or to constrain the data sets used for network training \citep{tagliabueEfficientDeepLearning2024}, employing neural projection layers to project control inputs into sets with specific properties \citep{paulsonApproximateClosedLoopRobust2020b, dontiEnforcingRobustControl2021a, chenApproximatingExplicitModel2018, maddalenaNeuralNetworkArchitecture2020}, giving probabilistic or statistical confidence levels \citep{kargEfficientRepresentationApproximation2020,hertneckLearningApproximateModel2018a,katzGeneratingProbabilisticSafety2023,zhangNearOptimalRapidMPC2021}, differentiable predictive control (DPC) with control barrier functions (CBFs) \citep{cortezDifferentiablePredictiveControl2022} or probabilistic guarantees \citep{drgonaLearningConstrainedParametric2024}, check and then correct methods \citep{hoseApproximateNonlinearModel2024}, specialized networks for local stability \citep{nakamura-zimmererNeuralNetworkOptimal2022}, leveraging the mixed-integer programming (MIP) representation or piecewise affine nature of certain networks \citep{fabianiReliablyStabilizingPiecewiseAffineNeural2023,schwanStabilityVerificationNeural2023, kargStabilityFeasibilityNeural2020, jouretSafetyVerificationNeuralNetworkBased2023}, and semidefinite programming (SDP) \citep{fazlyabSafetyVerificationRobustness2022}. Despite these efforts, there still lacks a fully interpretable network architecture that can provide mathematical guarantees with complete assurance and computational efficiency. 

Toward this direction, recent works have explored to design neural networks to approximate model predictive control policies. A given structure is considered for the control laws, typically continuous piecewise affine functions. Special NN architectures have been proposed to approximate these functions using neural projection layers \citep{chenApproximatingExplicitModel2018} and tunable quadratic programming layers \citep{maddalenaNeuralNetworkArchitecture2020}. It has also been shown that the solution to MPC problems can be represented by the difference of convex functions where each convex function is represented by an NN \citep{kargEfficientRepresentationApproximation2020}, or by using HardTanh-based NNs \citep{lupuExactRepresentationEfficient2024}. While these works provide guidance into the architecture necessary to represent these functions, it is not clear how the weights and biases of the networks can be realized without training. The reliance on training looses the continuously exact representation and therefore may fail to retain the mathematical guarantees (e.g., on stability). Furthermore, the weight and bias selections are highly dependent on the training data and may not generalize well to the entire state-space. Scalability also represents a challenge for the network training since every piece of the piecewise function must be learned and the number of pieces grows exponentially with the problem complexity. It is worth noting that, for linear systems, the optimal control laws theoretically determined by multi-parametric MPC (mp-MPC) are continuous piecewise affine functions which can be computed offline a priori \citep{bemporadExplicitLinearQuadratic2002a, pappasMultiparametricProgrammingProcess2021a}. This opens the opportunity to exactly reformulate the mp-MPC laws to neural networks without requiring any further training. Motivated by this, an exact representation is presented by Darup et al. \citep{darupExactRepresentationPiecewise2020} using rectified linear unit (ReLU) activation functions. However, this approach only applies to one-dimensional inputs which largely restricts its applicability in practice. 

In this work, we present Y-wise affine neural networks (YANNs), a fully explainable network architecture that exactly represents known piecewise affine functions without any training. YANNs can represent these functions defined on any amount of subdomains for any size of inputs and outputs. We utilize YANNs to exactly represent the explicit solution to mp-MPC problems. Since it is functionally equivalent to the optimal control policy, it inherits all control-theoretic properties such as recursive feasibility and stability. This functional equivalence is achieved by selecting the weights and biases of the network based on the parameters defining the piecewise affine function. We show that this approach can be extended to nonlinear functions where simple transformations exist to bring them to an affine form. We also show that YANNs evaluate quicker than traditional methods for piecewise affine functions regardless of if the amount of subdomains is in the tens, hundreds, or thousands. This speed-up can potentially push the boundaries for real-time control implementations.  

The remaining sections of this paper are organized as follows: Section 2 presents the necessary background of mp-MPC and establishes the nomenclature used in this paper. Section 3 gives the problem statement to be solved in this work. Section 4 proves the theoretical foundations of YANNs. Section 5 demonstrates the computational and practical advantages of YANNs through numerical case studies. Section 6 discusses the concluding remarks and our ongoing research directions. 
\section{Background and preliminaries}
\subsection{Multi-parametric model predictive control}
Consider the MPC regulation problem defined in Eq. \ref{eq:mpc}, which involves a linear time-invariant discrete-time system with equal operating and control horizons:
\begin{align}
\label{eq:mpc}
\min_u \quad & \sum_{k=1}^{N} \left( x_k^T Q x_k + u_k^T R u_k \right) + x_N^T P x_N \\
\text{s.t.} \quad & x_{k+1} = A x_k + B u_k, \quad \forall k \in \{1, 2, \cdots, N\} \nonumber \\
& y_{k} = C x_k + D u_k, \quad \forall k \in \{1, 2, \cdots, N\} \nonumber \\
& x_k \in \mathcal{X},y_k \in \mathcal{Y}, u_k \in \mathcal{U}, \quad \forall k \in \{1, 2, \cdots, N\} \nonumber \\
& x_N \in \mathcal{X}_f \nonumber
\end{align}
where $x_k \in \mathbb{R}^n$ is the system state at time step $k$, $u_k \in \mathbb{R}^m$ is the control input at time step $k$, $A$, $B$, $C$, and $D$ are system matrices, $Q$ and $R$ are symmetric positive semi-definite and positive definite weighting matrices, $P$ is the solution to the discrete-time algebraic Ricatti equation, $N \in \mathbb{N}$ is the prediction horizon, $\mathcal{X} = \{ x \in \mathbb{R}^n : F_x x \leq g_x \}$ is the state path constraint polytope, $\mathcal{Y} = \{ y \in \mathbb{R}^n : F_y y \leq g_y \}$ is the output constraint polytope, $\mathcal{U} = \{ u \in \mathbb{R}^m : F_u u \leq g_u \}$ is the input constraint polytope, $\mathcal{X}_f = \{ x \in \mathbb{R}^n : F_f x \leq g_f \}$ is the terminal set which is a control invariant polytope. 

This MPC regulation problem can be reformulated into an equivalent multi-parametric quadratic program (mp-QP), as given in Eq. \ref{eq:mp-QP mpc}:
\begin{align}
\label{eq:mp-QP mpc}
J^\ast(\theta) =\min_u \quad & u^THu +u^TZ\theta+\theta^T\hat{M}\theta \\
\text{s.t.} \quad & Gu\leq S\theta +W \nonumber \\
& CR_A\theta\leq CR_b \nonumber
\end{align}
where $\theta$ is the parametric set for mp-MPC. For this specific case, $\theta$ comprises the state $x$ at the current time step $k=1$. $u$ is the vector of $u_k,\text{ }\forall k\in[1,2,\cdots,N$] with $N$ being the prediction horizon from Eq. \ref{eq:mpc}. Weighting matrices $H$, $Z$, and $\hat{M}$ are formulated using the system matrices and MPC weighting matrices from Eq. \ref{eq:mpc}. Constraint matrices $G$, $S$, and $CR_A$, and constraint vectors $W$ and $CR_b$ are formulated using the system matrices and polytopic constraints from Eq. \ref{eq:mpc}. 

This equivalent mp-QP reformulation inherits the mathematical guarantees from the MPC problem such as recursive feasibility and stability. The solution to Eq. \ref{eq:mp-QP mpc} gives the optimal decision variables, $u^\ast$, as a continuous piecewise affine function of the uncertain parameters, $\theta$, that is defined on $p$ compact convex connected and non-overlapping polytopes. This is mathematically represented by Eq. \ref{eq:mp-QP soln}. 
\begin{equation}\label{eq:mp-QP soln}
u^\ast(\theta) =\begin{cases}
        K_1\theta + r_1,  \theta \in CR^1 = \{ CR_A^1\theta \leq CR_b^1\} \\
        \quad \vdots \\
        K_p\theta + r_p,  \theta \in CR^p = \{ CR_A^p\theta \leq CR_b^p\} \\
         \end{cases}
\end{equation}
where $K_i$ and $r_i$ are the coefficient matrix and constant vector for the explicit solution defined on polytope $i$, $\{CR_A^i\theta \leq CR_b^i\}$ defines the $i^{th}$ polytope $CR^i$, and $CR^i$ is a subset of the polytope formed by the $\theta$ constraints in Eq. \ref{eq:mp-QP mpc} such that $CR^i \subset \{CR_A\theta \leq CR_b\}$. These polytopic subdomains are conventionally referred to as critical regions in the context of multi-parametric programming. Similar reformulation strategies can be implemented for other types of MPC problems such as disturbance rejection, reference tracking, etc., where additional information is included in the $\theta$ vector \citep{sakizlisLinearModelPredictive2007}.

\subsection{Neural networks}
Consider a feed forward neural network with $L$ layers, input $X\in \mathbb{R}^n$, and output $Y^{[L]}\in\mathbb{R}^m$, as given in Eq. \ref{eq:NN}:
\begin{align}
    \label{eq:NN}
    &f(X) = Y^{[L]}\\
    &= \sigma^L(W^L(\sigma^{L-1}(W^{[L-1]}(\cdots\sigma^{[1]}(W^{[1]}X+B^{[1]})\nonumber\\
    &+B^{[L-1]})+B^L) \nonumber
\end{align}
where $W^{[i]}$, $B^{[i]}$, and $\sigma^{[i]}$ are the weights, bias, and activation function for layer $i$. The output of any layer $i$ is represented by $Y^{[i]}=\sigma^{[i]}(W^{[i]}Y^{[i-1]}+B^{[i]})$. For a fully connected network without skip or residual connections, $Y^{[i-1]}=X^{[i]}$. 

The types of activation functions of interest to this work are the ReLU, the binary step function (BSF), and the linear activation function as given in Eq. \ref{eq:ReLU}, Eq. \ref{eq:BSF}, and Eq. \ref{eq:linear} respectively. The use of these activation functions in YANNs will be discussed in detail in Section 4.
\begin{align}
    \label{eq:ReLU}
    &\text{ReLU}(x) = max(x,0)\\
    \label{eq:BSF}
    &\text{BSF}(x) = \begin{cases}
        1,& x\geq 0\\
        0,& \text{else}
    \end{cases}\\
    \label{eq:linear}
    &f(x)=x
\end{align}

\subsection{Further notation}
Let the indicator function for an input $x\in \mathbb{R}^n$ and a set $S$ be denoted by $\mathbbm{1}(x)$, where $\mathbbm{1}(x) = \begin{cases}
        1, & x\in S \\
        0, & \text{else}\end{cases}$. Additionally, we denote $a$ as a coefficient in a linear inequality or an affine equation. For a piecewise function with multiple inputs and outputs, $a_{i,j,k}$ is the coefficient for the $i^{th}$ input for the $j^{th}$ output for the $k^{th}$ subdomain. Similarly, we let $b$ denote the constant term for a linear inequality or affine equation. Lastly, we represent binary variables as $d$.
\section{Problem formulation}
Consider a continuous piecewise affine function, $f(x)=u$, with input $x\in \mathbb{R}^n$ and output $u\in \mathbb{R}^m$. As given in Eq. \ref{eq:affine func}, $f(x)$ is defined on $p$ compact convex polytopic subdomains, $S_k=Dom(f_k(x)) \; \forall k \in \{1, 2, \cdots, p\}$, that form a compact domain, $S=Dom(f(x))$. Correspondingly, there are $p$ subfunctions. The function $f(x)$ evaluates the $k^{th}$ subfunction $f_k(x)$ when $x$ is in the $k^{th}$ subdomain, $x\in S_k$. 
\begin{equation}\label{eq:affine func}
f(x) =\begin{cases}
        f_1(x) = u_1, &x\in S_1 \\
        \quad \vdots \\
         f_p(x)=u_p, &x\in S_p\\
         \end{cases}
\end{equation}
where $u_k$ is defined by $A_kx+b_k$ with $A$ and $b$ being the coefficient matrix and constant vector for the affine function defined on subdomain $k$. Each compact convex polytopic subdomain can be represented by the intersection of a system of linear inequalities through the half-space representation.

 The functions shown in Eq. \ref{eq:affine func} are the generalized form of the mp-MPC laws in Eq. \ref{eq:mp-QP soln}. The functions of this form also have wider applicability for the representation of piecewise linear process models and any decision-making solutions from multi-parametric linear/quadratic programming. The objective of this work is to exactly represent these functions through a neural network without needing any training. Thus, it is essential to show how the information from a known continuous piecewise affine function can be reformulated into a neural network. Motivated by the evaluation process of piecewise functions, we design logical network layers that efficiently identify the active subdomain by using the parameters within the linear constraints. This information is then used to evaluate the corresponding subfunction. The reformulation, subdomain identification, and function evaluation are presented in the proofs that follow. 
\section{Exact representation}
In this section, we present and prove the theoretical foundations of YANNs. There are two main components to the network: (i) constraint checking to determine the active subdomain, and (ii) function evaluation to provide the output of the corresponding subfunction. We start by proving simple cases and gradually build to the generalized forms. The full theorem will be presented at the end of this section. 
\subsection{Subdomain identification}
\begin{lem}
    \label{Lemma1}
    The indicator function of a linear inequality with an input $x\in \mathbb{R}^n$ can be expressed by a single neuron without any training. 
\end{lem}
\begin{pf*}{Proof.}
A linear inequality with input $x\in \mathbb{R}^n$ and its indicator function are defined by Eq. \ref{eq:LI} and Eq. \ref{eq:IF for 1 LI}, respectively.  
\begin{equation}
    \label{eq:LI}
    a_1x_1+a_2x_2+\cdots+a_nx_n\leq b
\end{equation}
\begin{equation}
    \label{eq:IF for 1 LI}
    \mathbbm{1}(x) = \begin{cases}
        1, & a_1x_1+a_2x_2+\cdots+a_nx_n\leq b \\
        0, & \text{else}
    \end{cases}
\end{equation}
Eq. \ref{eq:LI} can be rewritten as
\begin{equation}
    \label{eq:LI_negative}
    -a_1x_1-a_2x_2-\cdots-a_nx_n+b\geq0
\end{equation}
Applying the BSF to the left-hand side of Eq. \ref{eq:LI_negative} gives Eq. \ref{eq:Lem1 binary step}, which is equivalent to Eq. \ref{eq:IF for 1 LI}.
\begin{equation}
\label{eq:Lem1 binary step}
    BSF(x) = \begin{cases}
        1, & -a_1x_1-a_2x_2-\cdots-a_nx_n+b\geq0 \\
        0, & \text{else}
    \end{cases}
\end{equation}
Consider a neural network comprising a single node using the BSF as the activation function. The input is $X = [x_1, x_2,\cdots, x_n]^T$ and output is $Y$. Let the weights, $W$, be defined by the coefficients as represented in Eq. \ref{eq:LI_negative} such that $W = [-a_1,-a_2,\cdots, -a_n]$. Let the bias, $B$, be the constant in Eq. \ref{eq:LI_negative} such that $B = b$. Solving for $Y$ gives:
\begin{align*}
    Y & = \sigma(WX+B)  \\
    & = \sigma([-a_1,-a_2,\cdots -a_n]\times[x_1, x_2,\cdots x_n]^T+b)  \\
    & = \sigma(-a_1x_1-a_2x_2-\cdots-a_nx_n+b) \\
    & = \begin{cases}
        1, & -a_1x_1-a_2x_2-\cdots-a_nx_n+b\geq0 \\
        0, & \text{else}
        \end{cases}
\end{align*}
\begin{equation}
    \label{eq:Lem1 Y}
    \therefore Y = \mathbbm{1}(x)
\end{equation}
\end{pf*}
\begin{lem}
\label{Lemma2}
    The indicator function for a system of $q$ linear inequalities with input $x\in \mathbb{R}^n$ can be expressed by a two-layer neural network without any training. 
\end{lem}
\begin{pf*}{Proof.}
The system of inequalities and its indicator function are defined by Eq. \ref{eq:Lem2 Sys Ineq neg} and Eq. \ref{eq:Lem2 Ind F alt}, respectively, where $\mathbbm{1}_i(x)$ is the indicator function for inequality $i, \, \forall i\in\{1,2,\cdots,q\}$.
\begin{equation}
    \label{eq:Lem2 Sys Ineq neg}
    \begin{cases}
        -a_{1,1}x_{1}-a_{2,1}x_{2}-\cdots-a_{n,1}x_{n} + b_1 \geq0 \\
        -a_{1,2}x_{1}-a_{2,2}x_{2}-\cdots-a_{n,2}x_{n} + b_2 \geq0  \\
        \vdots\\
        -a_{1,q}x_{1}-a_{2,q}x_{2}-\cdots-a_{n,q}x_{n}+ b_q \geq0 \\
    \end{cases}
\end{equation}
\begin{equation}
    \label{eq:Lem2 Ind F alt}
    \mathbbm{1}(x) = \begin{cases}
        1, & \begin{cases}
        \mathbbm{1}_1(x) =1\\
        \mathbbm{1}_2(x) =1\\
        \vdots\\
        \mathbbm{1}_q(x) =1\\
    \end{cases} \\
        0, & \text{else}
    \end{cases}
\end{equation}
Applying the binary step function to the left-hand side of each inequality in the system represented by Eq. \ref{eq:Lem2 Sys Ineq neg} gives the following, which is equivalent to Eq. \ref{eq:Lem2 Ind F alt}.
\begin{equation*}
\label{eq:Lem2 binary step}
 BSF(x) = \begin{cases}
        1, & \begin{cases}
        -a_{1,1}x_{1}-a_{2,1}x_{2}-\cdots-a_{n,1}x_{n} + b_1 \geq0 \\
        -a_{1,2}x_{1}-a_{2,2}x_{2}-\cdots-a_{n,2}x_{n} + b_2 \geq0  \\
        \vdots\\
        -a_{1,q}x_{1}-a_{2,q}x_{2}-\cdots-a_{n,q}x_{n}+ b_q \geq0 \\
        \end{cases}\\
        0, & \text{else}
    \end{cases}
\end{equation*}
Consider a two-layer neural network with input $X = [x_1, x_2,\cdots, x_n]^T$. Layer one comprises $q$ nodes and layer two comprises one node. Use BSF as the activation function for the first layer and ReLU as the activation function for the second layer. Let the weights and biases of layer one, $W^{[1]} \in \mathbb{R}^{q\times n}$ and $B^{[1]}\in \mathbb{R}^{q}$, be defined by the coefficients and constants in Eq. \ref{eq:Lem2 Sys Ineq neg}, as shown below: 
\begin{equation*}
    W^{[1]} = \begin{bmatrix}
        -a_{1,1} & -a_{2,1} & \cdots & -a_{n,1}\\
        -a_{1,2} & -a_{2,2} & \cdots & -a_{n,2}\\
        \vdots & \vdots & \vdots & \vdots\\
        -a_{1,q} & -a_{2,q} & \cdots & -a_{n,q}\\
    \end{bmatrix}
    , B^{[1]} = \begin{bmatrix}
        b_1\\
        b_2\\
        \vdots\\
        b_q
    \end{bmatrix}
\end{equation*}
Solving for $Y^{[1]}$ gives:
\begin{align*}
    Y^{[1]} & = \sigma^{[1]}(W^{[1]}X+B^{[1]})  \\
    & = \sigma^{[1]}(\begin{bmatrix}
        -a_{1,1} & -a_{2,1} & \cdots & -a_{n,1}\\
        -a_{1,2} & -a_{2,2} & \cdots & -a_{n,2}\\
        \vdots & \vdots & \vdots & \vdots\\
        -a_{1,q} & -a_{2,q} & \cdots & -a_{n,q}\\
    \end{bmatrix} \times \begin{bmatrix}
        x_1\\
        x_2\\
        \vdots\\
        x_n
    \end{bmatrix}+\begin{bmatrix}
        b_1\\
        b_2\\
        \vdots\\
        b_q
    \end{bmatrix}) \\
    & = \sigma^{[1]}(\begin{bmatrix}
        -a_{1,1}x_1-a_{2,1}x_2-\cdots-a_{n,1}x_n+b_1\\
        -a_{1,2}x_1 -a_{2,2}x_2-\cdots-a_{n,2}x_n+b_2\\
        \vdots\\
        -a_{1,q}x_1-a_{2,q}x_2-\cdots-a_{n,q}x_n+b_q\\
    \end{bmatrix}\\
    &= \begin{bmatrix}
        \begin{cases}1,-a_{1,1}x_1-a_{2,1}x_2-\cdots-a_{n,1}x_n+b_1 \geq0\\
        0, \text{ else}
        \end{cases}\\
        \begin{cases}
            1, -a_{1,2}x_1 -a_{2,2}x_2-\cdots-a_{n,2}x_n+b_2\geq0\\
            0, \text{ else}
        \end{cases}\\
        \vdots\\
        \begin{cases}
            1, -a_{1,q}x_1-a_{2,q}x_2-\cdots-a_{n,q}x_n+b_q\geq0\\
            0, \text{ else}
        \end{cases}\\
    \end{bmatrix}
\end{align*}
Let the weights of layer two be the $1$-vector of size $q$, $W^{[2]} = [1,1,\cdots,1] \in \mathbb{R}^{q}$, and the bias be defined by, $B^{[2]}=-q+1$. Solving for $Y^{[2]}$ gives:
\begin{align*}
    Y^{[2]} &= \sigma^{[2]}(W^{[2]}Y^{[1]}+B^{[2]})\\
    & = \sigma^{[2]}(\sum_{l=1}^qY^{[1]}-q+1)\\
    &=\max(\sum_{l=1}^qY^{[1]}-q+1,0)\\
    &=\begin{cases}
        1,&\sum_{l=1}^qY^{[1]}=q\\
        0, &\text{else}
    \end{cases}
\end{align*}
The expression $\sum_{l=1}^qY^{[1]}$ can at most evaluate to $q$ since $Y^{[1]}$ is a vector of binaries with dimension $q$. The only case where $\sum_{l=1}^qY^{[1]}=q$ is when all the $q$ inequalities are satisfied.
\begin{equation}
    \label{eq:Lem2 Y}
    \therefore Y^{[2]} = \mathbbm{1}(x)
\end{equation}
\end{pf*}
\begin{rem}
    The results of Lemma \ref{Lemma2} provide a network that efficiently checks a system of linear inequality constraints in a simultaneous manner. Equality constraints can also be checked by including both the negative and positive forms of the inequality in the network. Furthermore, if there exists a transformation to bring nonlinear inequality constraints into a linear form, these constraints can also be checked via this network. Example 1 showcases the extensions using an instance of a nonlinear equation. 
\end{rem}
\begin{pf*}{Example 1}
    Consider the ellipsoid defined by $2(x-2)^2+y^2+z^2=5$. This equation can be rewritten to the form $2x^2+y^2+z^2-8x+3=0$. To determine if a point lies on this ellipsoid, the following two inequalities must be simultaneously true, i.e. $2x^2+y^2+z^2-8x+3\geq0$ and $-1(2x^2+y^2+z^2-8x+3)\geq0$. This can be achieved with a network of the form presented in Lemma \ref{Lemma2}. This network can use an input of $X=[x^2, y^2, z^2, x]$ with $W^{[1]}=[2,1,1,-8; -2,-1, -1, 8]$, $B^{[1]}=[3,-3]^T$, $W^{[2]}=[1,1]$, and $B^{[2]}=-1$. The output of this network, $Y^{[2]}$, is one if both inequalities are true, implying the input is on the ellipsoid, and is zero otherwise. A smaller network can be used if the network input is defined as $X=[(x-2)^2, y^2,z^2]$.
\end{pf*}
\begin{lem}
\label{Lemma4}
    The solutions of indicator functions for $p$ systems of linear inequalities can be represented via a two-layer neural network without any training, using $x\in \mathbb{R}^n$ as input and a vector of size $p$ as output.
\end{lem}
\vspace{-0.5cm}
\begin{pf*}{Proof.}Let $q_{s}$ be the amount of inequalities in the $s^{th}$ system and let $q = \sum_{s=1}^pq_s$ which represents the total amount of inequalities across all $p$ systems. The $s^{th}$ system of inequalities can be written as:
\begin{equation}
\label{eq:lem3 system s}
    \begin{cases}
        -a_{1,1,s}x_{1}-a_{2,1,s}x_{2}-\cdots-a_{n,1,s}x_{n} + b_{1,s} \geq0 \\
        -a_{1,2,s}x_{1}-a_{2,2,s}x_{2}-\cdots-a_{n,2,s}x_{n} + b_{2,s} \geq0  \\
        \vdots\\
        -a_{1,q_s,s}x_{1}-a_{2,q_s,s}x_{2}-\cdots-a_{n,q_s,s}x_{n}+ b_{q_s,s} \geq0 \\
    \end{cases}
\end{equation}
Similar to Eq. \ref{eq:Lem2 Ind F alt}, an indicator function can be written for each system of inequalities where $\mathbbm{1}_s(x)$ is the indicator function for the $s^{th}$ system. Consider a two-layer neural network with input $X = [x_1, x_2,\cdots, x_n]^T$. Layer one consists of $q$ nodes and layer two consists of $p$ nodes. The activation functions are $\sigma^{[1]}=\text{BSF}(x)$ and $\sigma^{[2]}=\text{ReLU}(x)$, with weights and biases defined below.
\begin{equation*}
    W^{[1]} = \begin{bmatrix}
        -a_{1,1,1} & \cdots & -a_{n,1,1}\\
        \vdots & \vdots & \vdots\\
        -a_{1,q_1,1} & \cdots & -a_{n,q_1,1}\\
        -a_{1,1,2} & \cdots & -a_{n,1,2}\\
        \vdots & \vdots & \vdots\\
        -a_{1,q_p,p} & \cdots & -a_{n,q_p,p}\\
    \end{bmatrix}
    , B^{[1]} = \begin{bmatrix}
        b_{1,1}\\
        \vdots\\
        b_{q_1,1}\\
        b_{1,2}\\
        \vdots\\
        b_{q_p,p}
    \end{bmatrix}
\end{equation*}
\begin{equation*}
    W^{[2]} = \begin{bmatrix}
        [1]_{q_1} & 0 &\cdots & 0\\
        [0]_{q_1} & [1]_{q_2}& \vdots& 0\\
        \vdots & \ddots & \ddots&\vdots\\
        0 & 0 &\cdots&[1]_{q_p}\\
    \end{bmatrix}
    , B^{[2]} = \begin{bmatrix}
        1-q_1\\
        1-q_2\\
        \vdots\\
         1-q_p
    \end{bmatrix}
\end{equation*}
\begin{align*}
    &\text{where }W^{[1]}\in\mathbb{R}^{q\times n},B^{[1]}\in\mathbb{R}^{q},W^{[2]}\in\mathbb{R}^{p \times q},B^{[2]}\in\mathbb{R}^{p} \\
\end{align*}
Solving for $Y^{[1]}$ gives:
\begin{small}
\begin{align*}
    &Y^{[1]} = \sigma^{[1]}(W^{[1]}X+B^{[1]}) =\\
    &\begin{bmatrix*}[l]
        \begin{cases}1,-a_{1,1,1}x_1-a_{2,1,1}x_2-\cdots-a_{n,1,1}x_n+b_{1,1} \geq0\\
        0, \text{else}
        \end{cases}\\
        \multicolumn{1}{c}{\vdots}\\
        \begin{cases}
            1, -a_{1,q_1,1}x_1 -a_{2,q_1,1}x_2-\cdots-a_{n,q_1,1}x_n+b_{q_1,1}\geq0\\
            0, \text{else}
        \end{cases}\\
        \multicolumn{1}{c}{\vdots}\\
        \begin{cases}
            1, -a_{1,1,2}x_1 -a_{2,1,2}x_2-\cdots-a_{n,1,2}x_n+b_{1,2}\geq0\\
            0, \text{else}
        \end{cases}\\
        \multicolumn{1}{c}{\vdots}\\
        \begin{cases}
            1, -a_{1,q_p,p}x_1-a_{2,q_p,p}x_2-\cdots-a_{n,q_p,p}x_n+b_{q_p,p}\geq0\\
            0, \text{else}
        \end{cases}\\
    \end{bmatrix*}
\end{align*}
\end{small}

Solving for $Y^{[2]}$ gives:
\begin{align*}
    Y^{[2]} &= \sigma^{[2]}(W^{[2]}Y^{[1]}+B^{[2]})\\
    & = \sigma^{[2]}(\begin{bmatrix*}[l]
        \sum_{l=1}^{q_1}Y^{[1]}_l-q_1+1\\
        \sum_{l=q_1+1}^{q_2+q_1}Y^{[1]}_l-q_2+1\\
        \sum_{l=q_2+q_1+1}^{q_3+q_2+q_1}Y^{[1]}_l-q_3+1\\
        \multicolumn{1}{c}{\vdots}\\
        \sum_{l=q-q_p}^{q}Y^{[1]}_l-q_p+1
    \end{bmatrix*})\\
    &=\begin{bmatrix*}[l]
        \max(\sum_{l=1}^{q_1}Y^{[1]}_l-q_1+1,0)\\
        \max(\sum_{l=q_1+1}^{q_2+q_1}Y^{[1]}_l-q_2+1,0)\\
        \max(\sum_{l=q_2+q_1+1}^{q_3+q_2+q_1}Y^{[1]}_l-q_3+1,0)\\
        \multicolumn{1}{c}{\vdots}\\
        \max(\sum_{l=q-q_p}^{q}Y^{[1]}_l-q_p+1,0)
    \end{bmatrix*}\\
    &=\begin{bmatrix*}[l]
        \begin{cases}
        1,&\sum_{l=1}^{q_1}Y^{[1]}_l=q_1\\
        0, &\text{else}
    \end{cases}\\
    \begin{cases}
        1,&\sum_{l=q_1+1}^{q_2+q_1}Y^{[1]}_l=q_2\\
        0, &\text{else}
    \end{cases}\\
    \multicolumn{1}{c}{\vdots}\\
    \begin{cases}
        1,&\sum_{l=q-q_p}^{q}Y^{[1]}_l=q_p\\
        0, &\text{else}
    \end{cases}
    \end{bmatrix*}
\end{align*}
For a system $s$ where $s\in\{1, 2, \cdots, p\}$, the only case where $\sum_{l=1+q_1+\cdots+q_{s-1}}^{q_1+\cdots+q_{s}}Y^{[1]}_l=q_s$ is when all the inequalities in this system are satisfied.
\begin{equation}
    \label{eq:lem3 Y final}
    \therefore Y^{[2]} = \begin{bmatrix}
        \mathbbm{1}_1(x)\\
        \mathbbm{1}_2(x)\\
        \vdots\\
        \mathbbm{1}_p(x)
    \end{bmatrix}
\end{equation}
\end{pf*}
\begin{rem}The result of Lemma \ref{Lemma4} is useful in identifying which systems of linear inequalities are satisfied while also providing an efficient way to do so. However, it is possible that there exist more than one elements in the resulting $Y^{[2]}$ vector taking the value of $1$ while the other elements are $0$. For example, if the inputs lie on the boundary between two adjacent subdomains, the $Y^{[2]}$ vector will have two elements with the value of $1$ as the inputs simultaneously satisfy the system of inequalities defining both subdomains. For the purpose of YANNs, it is essential to force that only one of these elements takes the value of $1$, e.g. $f([0, 1, 1, 0]^T) = [0,1,0,0]^T$. A solution to achieve this is provided in the following Lemma. 

\end{rem}
\begin{lem}
\label{Lemma6}
    Given an input vector of binary variables in the size of $s$, it is possible to use a single-layer neural network to return an output vector of binary variables in the same size. All the values in the output vector are $0$ except for one element as $1$, the index of which corresponds to the index of the first binary with value $1$ in the input vector, if there are any.
\end{lem}
\begin{pf*}{Proof.}
Consider a single-layer neural network with $p$ nodes. The input vector is $X = [d_1, d_2, \cdots, d_s]^T$ of binary variables and the output vector is $Y$ of binary variables. ReLU is used as the activation function with the weights, $W\in \mathbb{R}^{s\times s}$, and bias, $B\in \mathbb{R}^{s}$, defined below.
\begin{equation*}
    W = \begin{bmatrix}
        1 & 0 & 0 & 0 &\cdots & 0 & 0\\
        -1 &1 & 0 & 0 & \cdots & 0 & 0\\
        -1 & -1 & 1 &0 & \cdots & 0 &0\\
        \vdots & \vdots & \ddots & \ddots &\ddots & \vdots & \vdots\\
        -1 & -1& -1 & -1 & \cdots & 1 & 0\\
        -1 & -1 & -1 & -1 & \cdots & -1 & 1
    \end{bmatrix}
    , B = 0
\end{equation*}
Solving for Y gives:
\begin{align*}
    Y& =\sigma(WX+B)  \\
    &= \sigma(\begin{bmatrix}
        1 & 0 & 0 & 0 &\cdots & 0 & 0\\
        -1 &1 & 0 & 0 & \cdots & 0 & 0\\
        -1 & -1 & 1 &0 & \cdots & 0 &0\\
        \vdots & \vdots & \ddots & \ddots &\ddots & \vdots & \vdots\\
        -1 & -1& -1 & -1 & \cdots & 1 & 0\\
        -1 & -1 & -1 & -1 & \cdots & -1 & 1
    \end{bmatrix}\times \begin{bmatrix}
        d_1\\
        d_2\\
        d_3\\
        \vdots\\
        d_{s-1}\\
        d_{s}
    \end{bmatrix})\\
    &=\begin{bmatrix*}[l]
        \max(d_1,0)\\
        \max (d_2 - d_1, 0)\\
        \max (d_3 - (d_2+d_1), 0)\\
        \multicolumn{1}{c}{\vdots}\\
        \max (d_s - \sum_{l=1}^{s-1}d_l,0)
    \end{bmatrix*}
\end{align*}
where the $i^{th}$ element of $Y$ is given by
\begin{equation}
    \label{lem6 yi}
    Y_i = \max (d_i - \sum_{l=1}^{i-1}d_l, 0)
\end{equation}
Three cases may exist to further analyze from Eq. \ref{lem6 yi}: (i) none of the binaries in the input vector $X$ take the value of 1, (ii) there is exactly one binary in $X$ taking the value of 1, and (iii) there are more than one binaries in $X$ taking the value of 1.

\textbf{Case i:} Assume there is no $d_i\in X$ such that $d_i = 1$. The following can be resolved from Eq. \ref{lem6 yi}:
\begin{align*}
    Y_i & = \max (d_i - \sum_{l=1}^{i-1}d_l, 0) = \max (0-0,0) = 0\\
    \implies Y_i & = 0 \quad \forall i \in \{1, 2, \cdots, s\}\\
    \implies Y &= [0]_s
\end{align*}
\textbf{Case ii:} Assume there is only one element in $X$ with index $i$ such that $d_i = 1$. Let $h$ be an index such that $h<i$ and $j$ be an index such that $i<j$. The following can be resolved from Eq. \ref{lem6 yi}:
\begin{align*}
    Y_h & = \max (d_h - \sum_{l=1}^{h-1}d_l, 0) = \max (0-0,0) = 0\\
    \implies Y_h &= 0 \quad \forall h\in\{1, 2, \cdots, i-1\}\\
    Y_i & = \max (d_i - \sum_{l=1}^{i-1}d_l, 0) = \max (1-0,0) = 1\\
    Y_j & = \max (d_j - \sum_{l=1}^{j-1}d_l, 0) = \max (0-1,0) = 0\\
    \implies Y_j &= 0 \quad \forall j\in\{i+1, i+2, \cdots, s\}\\
    \implies Y^T &= \begin{bmatrix}
        0 & \cdots & 0 & 1 & 0 &\cdots& 0
    \end{bmatrix}, \sum Y=1, Y_i=1
\end{align*}
\textbf{Case iii:} Assume there are at least two elements in $X$ taking the value of 1. $i$ and $j$ are the indexes of the first and second element of $X$ with value one respectively, such that $d_i = 1, d_j = 1$. Note that $i<j$. Let $h$ be an index such that $h<i$, let $g$ be an index such that $i<g<j$, and let $k$ be an index such that $j<k$. Let the variable $v = \sum_{l=1}^{k-1}d_l$ and the variable $t = d_k$. The following can be resolved From Eq. \ref{lem6 yi}:
\begin{align*}
    Y_h & = \max (d_h - \sum_{l=1}^{h-1}d_l, 0) = \max (0-0,0) = 0\\
\implies Y_h &= 0 \quad \forall h\in\{1, 2, \cdots, i-1\}\\
    Y_i & = \max (d_i - \sum_{l=1}^{i-1}d_l, 0) = \max (1-0,0) = 1\\
    Y_g & = \max (d_g - \sum_{l=1}^{g-1}d_l, 0) = \max (0-1,0) = 0\\
    Y_j & = \max (d_j - \sum_{l=1}^{j-1}d_l, 0) = \max (1-1,0) = 0\\
    v &=\sum_{l=1}^{k-1}d_l \geq 2, \quad t = d_k \in \{0,1\}\\
    Y_k & = \max (d_k - \sum_{l=1}^{k-1}d_l, 0) = \max (t-v,0) = 0\\
    \implies Y_l &= 0 \quad \forall l\in\{i+1, i+2, \cdots, s\}\\
    \implies Y^T &= \begin{bmatrix}
        0 & \cdots & 0 & 1 & 0 &\cdots& 0
    \end{bmatrix}, \sum Y=1, Y_i=1
\end{align*}
The above three cases cover all possibilities for a vector of binary variables of dimension $s$. Therefore, the network presented sufficiently proves Lemma \ref{Lemma6}.
\end{pf*}
\begin{rem}
    If it is desired to distinctly identify the case when all the binaries in the input vector $X$ are zero, an additional node can be added to the network layer used in this proof with a weighting vector where all elements are negative one and a bias of one. 
\end{rem}

\subsection{Function Evaluation}
\begin{lem}
\label{Lemma8}
    Any affine equation with input $x\in \mathbb{R}^n$ can be exactly represented by a two-layer neural network without any training. 
\end{lem}
\begin{pf*}{Proof.}
An n-dimensional affine equation can be written in two equivalent ways as given in Eq. \ref{eq:lem7 affine pos} and Eq. \ref{eq:lem7 affine neg}.
\begin{equation}
    \label{eq:lem7 affine pos}
    a_1x_1+a_2x_2+\cdots+a_nx_n+b=u
\end{equation}
\begin{equation}
    \label{eq:lem7 affine neg}
    -a_1x_1-a_2x_2-\cdots-a_nx_n-b=-u
\end{equation}
Consider a two-layer neural network with input $X = [x_1, x_2, \cdots, x_n]^T$. The first layer consists of two nodes and the second layer consists of one node. The activation functions are $\sigma^{[1]} = \text{ReLU}(x)$ and $\sigma^{[2]} = x$, with the weights and biases defined below:
\begin{align*}
    W^{[1]} = \begin{bmatrix}
        a_1 & a_2 & \cdots & a_n\\
        -a_1 & -a_2 & \cdots & -a_n
    \end{bmatrix} &, B^{[1]} = \begin{bmatrix}
        b\\-b
    \end{bmatrix}\\
    W^{[2]} = \begin{bmatrix}
        1 &-1
    \end{bmatrix} &, B^{[2]} = 0
\end{align*}
Solving for $Y^{[1]}$ gives:
\begin{align*}
    Y^{[1]} &= \sigma^{[1]}(W^{[1]}X+B^{[1]})\\
    &= \sigma^{[1]}(\begin{bmatrix}
        a_1 & a_2 & \cdots & a_n\\
        -a_1 & -a_2 & \cdots & -a_n
    \end{bmatrix}\times\begin{bmatrix}
        x_1\\x_2\\ \vdots \\ x_n
    \end{bmatrix} + \begin{bmatrix}
        b\\-b
    \end{bmatrix})\\
    &=\begin{bmatrix*}[l]
        \max (u, 0)\\ \max(-u,0)
    \end{bmatrix*}\\
\end{align*}
Solving for $Y^{[2]}$ gives:
\begin{align*}
    Y^{[2]} &= \sigma^{[2]}(W^{[2]}Y^{[1]}+B^{[2]})\\
    &=W^{[2]}Y^{[1]}\\
    &= max(u,0)-max(-u,0)\\
    &= \begin{cases}
        u, & u>0\\
        u, &u<0\\
        0, &u=0\\
    \end{cases}\\
\end{align*}
\vspace{-0.5cm}
\begin{equation}
    \label{eq:lem7 y2 final}
    \therefore Y^{[2]} = u
\end{equation}
\end{pf*}
\begin{rem}
    It is possible to represent certain nonlinear equations using the same approach that was presented in Example 1, by leveraging nonlinear inputs to transform the nonlinear equality to a linear equality.
\end{rem}
\begin{lem}
\label{Lemma 12}
    Any system of affine functions with input $x\in \mathbb{R}^n$ and output $u\in \mathbb{R}^m$ can be represented by a two-layer neural network without any training. 
\end{lem}
\vspace{-0.5cm}
\begin{pf*}{Proof.}
An n-dimensional system of affine functions, $f(x) = u$, $u\in \mathbb{R}^m$, can be written as a system of equations in the two equivalent ways provided in Eq. \ref{eq:lem11 affine pos} and Eq. \ref{eq:lem11 affine neg}.
\begin{equation}
    \label{eq:lem11 affine pos}
    \begin{cases}
        a_{1,1}x_1+a_{2,1}x_2+\cdots+a_{n,1}x_n+b_1=u_1\\
        a_{1,2}x_1+a_{2,2}x_2+\cdots+a_{n,2}x_n+b_2=u_2\\
        \multicolumn{1}{c}{\vdots}\\
        a_{1,m}x_1+a_{2,m}x_2+\cdots+a_{n,m}x_n+b_m=u_m\\
    \end{cases}
\end{equation}
\begin{equation}
    \label{eq:lem11 affine neg}
    \begin{cases}
        -a_{1,1}x_1-a_{2,1}x_2-\cdots-a_{n,1}x_n-b_1=-u_1\\
        -a_{1,2}x_1-a_{2,2}x_2-\cdots-a_{n,2}x_n-b_2=-u_2\\
        \multicolumn{1}{c}{\vdots}\\
        -a_{,1m}x_1-a_{2,m}x_2-\cdots-a_{n,m}x_n-b_m=-u_m\\
    \end{cases}
\end{equation}
Consider a two-layer neural network with input $X = [x_1, x_2, \cdots, x_n]^T$. The first layer consists of $2m$ nodes and the second layer consists of $m$ nodes. The activation functions are $\sigma^{[1]} = \text{ReLU}(x)$ and $\sigma^{[2]} = x$, with the weights and biases defined below. Note that this architecture consists of $m$ independent sub-networks of the type presented in Lemma \ref{Lemma8}.
\begin{align*}
    &W^{[1]} = \begin{bmatrix}
        a_{1,1} & a_{2,1} & \cdots & a_{n,1}\\
        -a_{1,1} & -a_{2,1} & \cdots & -a_{n,1}\\
        \vdots&\vdots&\vdots&\vdots\\
        a_{1,m} & a_{2,m} & \cdots & a_{n,m}\\
        -a_{1,m} & -a_{2,m} & \cdots & -a_{n,m}\\
    \end{bmatrix} , B^{[1]} = \begin{bmatrix}
        b_1\\-b_1\\ \vdots\\ b_m\\-b_m
    \end{bmatrix}\\
    &W^{[2]} = \begin{bmatrix}
        1 &-1 & 0 & 0 & \cdots & 0 & 0\\
        0 &0 & 1 & -1 & \cdots & 0 & 0\\
        \vdots&\vdots&\ddots&\ddots&\ddots&\ddots&\ddots\\
        0 &0 & 0 & 0 & \cdots & 1 & -1
    \end{bmatrix} \in \mathbb{R}^{m\times2m} , B^{[2]} = 0
\end{align*}
Solving for $Y^{[1]}$ gives:
\begin{align*}
    Y^{[1]} &= \sigma^{[1]}(W^{[1]}X+B^{[1]})\\
    &=\begin{bmatrix*}[l]
        \max (u_1, 0)\\ \max(-u_1,0) \\ \multicolumn{1}{c}{\vdots} \\ \max (u_m, 0)\\ \max(-u_m,0)
    \end{bmatrix*}\\
\end{align*}
Solving for $Y^{[2]}$ gives:
\begin{align*}
    Y^{[2]} &= \sigma^{[2]}(W^{[2]}Y^{[1]}+B^{[2]})\\
    &=W^{[2]}Y^{[1]}\\
    &= \begin{bmatrix*}[l]
        max(u_1,0)-max(-u_1,0)\\
        \multicolumn{1}{c}{\vdots}\\
        max(u_m,0)-max(-u_m,0)\\
    \end{bmatrix*}\\
    &= \begin{bmatrix*}[l]
            \begin{cases}
            u_1, & u_1>0\\
            u_1, &u_1<0\\
            0, &u_1=0\\
        \end{cases}\\
        \multicolumn{1}{c}{\vdots}\\
        \begin{cases}
            u_m, & u_m>0\\
            u_m, &u_m<0\\
            0, &u=0\\
        \end{cases}\\
    \end{bmatrix*}
\end{align*}
\begin{equation}
    \label{eq:lem11 y2 final}
    \therefore Y^{[2]} = u
\end{equation}
\end{pf*}
\begin{rem}
    The previous two Lemmas show that ReLU-based networks can represent the outputs from a system of affine equations of any dimensions. In what follows, we will extend the ReLU representation to piecewise affine function over various input subdomains and constrain certain outputs to zero when the corresponding subdomain is inactive.
\end{rem}
\begin{lem}
\label{Lemma 13}
    For a given piecewise affine function, $f(x)=u$ with input $x\in\mathbb{R}^n$ and output $u\in \mathbb{R}^m$ that is defined on $p$ compact subdomains, $Dom(f_k(x)), \, \forall k \in \{1, 2, \cdots, p\}$, which form a compact domain, $Dom(f(x))$, there exists a big constant number, $M$, such that $M\geq |f_k(x)|, \,\forall k \in \{1, 2, \cdots, p\},\,\forall x\in Dom(f(x))$. 
\end{lem}
\vspace{-.5cm}
\begin{pf*}{Proof.}
Each $f_k(x),\, \forall k \in \{1, 2, \cdots, p\}$ is a continuous affine function that can be defined on all $x\in \mathbb{R}^n$ which implies that it can be defined on any subset of $\mathbb{R}^n$. Since the full domain of the piecewise affine function, $Dom(f(x))$, is compact and contained within $\mathbb{R}^n$, then by the bounded function theorem (Appendix \ref{Bounded function th}), there exists a number $M$ such that $M\geq |f_k(x)|,\, \forall k \in \{1, 2, \cdots, p\},\, \forall x\in Dom(f(x))$.
\end{pf*}
\begin{rem}
    In the case of YANNs, it is essential to choose the big $M$ value to be larger than the value of any subfunction $f_k(x)$ over the full domain $x\in Dom(f(x))$ instead of its corresponding subdomain $x\in Dom(f_k(x))$. The rationale to this is elucidated through the proof of the following Lemma.
\end{rem}
 \begin{lem}
 \label{Lemma 15}
     Given an affine equation with input $x\in \mathbb{R}^n$ which is defined on a compact set, $S$. Using the indicator function as an additional input, this equation can be expressed by a two-layer neural network without any training for all $x\in S$.
 \end{lem}
 \vspace{-.5cm}
 \begin{pf*}{Proof.}
An n-dimensional affine equation can be written in two equivalent ways as given previously in Eq. \ref{eq:lem7 affine pos} and Eq. \ref{eq:lem7 affine neg}. Consider a two-layer neural network with input $X = [\mathbbm{1}(x), x_1, x_2, \cdots, x_n]^T$. The first layer comprises two nodes and the second layer comprises one node. The activation functions are $\sigma^{[1]} = \text{ReLU}(x)$ and $\sigma^{[2]} = x$. The weights and biases are defined below. Note that a key difference between this network and the one presented in the proof of Lemma \ref{Lemma8} is that the indictor function $\mathbbm{1}(x)$ is used as an additional input. 
\begin{align*}
    W^{[1]} = \begin{bmatrix}
        M& a_1 & a_2 & \cdots & a_n\\
        M& -a_1 & -a_2 & \cdots & -a_n
    \end{bmatrix} &, B^{[1]} = -M+\begin{bmatrix}
        b\\-b
    \end{bmatrix}\\
    W^{[2]} = \begin{bmatrix}
        1 &-1
    \end{bmatrix} &, B^{[2]} = 0\\
    \text{where} \quad M \geq |u| \quad \forall x\in S
\end{align*}
Solving for $Y^{[1]}$ gives:
\begin{align*}
    Y^{[1]} &= \sigma^{[1]}(W^{[1]}X+B^{[1]})\\
    &= \sigma^{[1]}(\begin{bmatrix}
        M& a_1 & a_2 & \cdots & a_n\\
        M& -a_1 & -a_2 & \cdots & -a_n
    \end{bmatrix}\times\begin{bmatrix}
        \mathbbm{1}(x)\\x_1\\x_2\\ \vdots \\ x_n
    \end{bmatrix}\\& + \begin{bmatrix}
        b-M\\-b-M
    \end{bmatrix})\\
    &=\begin{bmatrix}
        \max (M\mathbbm{1}(x)+u-M, 0)\\ \max(M\mathbbm{1}(x)-u-M,0)
    \end{bmatrix}\\
\end{align*}
Evaluating $\mathbbm{1}(x)$ and solving for $Y^{[2]}$ gives:
\begin{align*}
    x\in S&\implies\mathbbm{1}(x) =1\\
    Y^{[2]} &= \sigma^{[2]}(W^{[2]}Y^{[1]}+B^{[2]})\\
    &=W^{[2]}Y^{[1]}\\
    = max(M\mathbbm{1}(x)&+u-M,0)-max(M\mathbbm{1}(x)-u-M,0)\\
    &= max(u,0)-max(-u,0)\\
    &= \begin{cases}
        u, & u>0\\
        u, &u<0\\
        0, &u=0\\
    \end{cases}\\
\end{align*}
\vspace{-1cm}
\begin{equation}
    \therefore Y^{[2]} = u : x\in S
\end{equation}
\end{pf*}
\begin{rem}
    It can also be proven that the output of the network, $Y^{[2]}$, evaluates to $0$ when $x\notin S$ as long as $M\geq |u|$. Lemma \ref{Lemma 15} can thus be leveraged to switch the network between evaluating the affine function when $x\in S$ or outputting $0$ when $x\notin S$. This provides the key underpinning idea in what follows. 
\end{rem}
\begin{lem}
\label{Lemma 17}
    Given any continuous piecewise affine function, $f(x)=u$, with input $x\in \mathbb{R}^n$ and output $u\in \mathbb{R}^m$. This function is defined on $p$ compact subdomains, $S_k=Dom(f_k(x)), \, \forall k \in \{1, 2, \cdots, p\}$, that form a compact domain, $S=Dom(f(x))$. It can be expressed by a two-layer neural network without any training, given the indicator functions for each subdomain and assuming the input $x$ is contained within one and only one subdomain.
\end{lem}
\vspace{-0.5cm}
\begin{pf*}{Proof.}
 Given a piecewise affine function, $f(x)=u$, with input $x$ of dimension $n$ and output $u$ of dimension $m$. Assume that this function is defined on $p$ compact subdomains, $S_k=Dom(f_k(x)), \, \forall k \in \{1, 2, \cdots, p\}$. The $k^{th}$ system of equations can be written in the two following equivalent forms given in Eq. \ref{lem16 pos} and Eq. \ref{lem16 neg}. 
 
\begin{equation}
\label{lem16 pos}
    \begin{cases}
        a_{1,1,k}x_1+\cdots+a_{n,1,k}x_n+b_{1,k}=u_{1,k}\\
        a_{1,2,k}x_1+\cdots+a_{n,2,k}x_n+b_{2,k}=u_{2,k}\\
        \multicolumn{1}{c}{\vdots}\\
        a_{1,m,k}x_1+\cdots+a_{n,m,k}x_n+b_{m,k}=u_{m,k}\\
    \end{cases}
\end{equation}
\begin{equation}
\label{lem16 neg}
    \begin{cases}
        -a_{1,1,k}x_1-\cdots-a_{n,1,k}x_n-b_{1,k}=-u_{1,k}\\
        -a_{1,2,k}x_1-\cdots-a_{n,2,k}x_n-b_{2,k}=-u_{2,k}\\
        \multicolumn{1}{c}{\vdots}\\
        -a_{1,m,k}x_1-\cdots-a_{n,m,k}x_n-b_{m,k}=-u_{m,k}\\
    \end{cases}
\end{equation}
Consider a two-layer neural network with input $X = [\mathbbm{1}_1(x), \mathbbm{1}_2(x), \cdots, \mathbbm{1}_k(x), x_1, x_2, \cdots, x_n]^T$. The first layer comprises $2mp$ nodes and the second layer comprises $m$ nodes. The activation functions are $\sigma^{[1]} = \text{ReLU}(x)$ and $\sigma^{[2]} = x$. The weights and biases are defined below.
\begin{align*}
    &W^{[1]T} =\\
    &\begin{bNiceArray}{ccccccc}
        [M]_{2m} & 0 &\cdots&\cdots&\cdots & 0 & 0\\
        [0]_{2m} & [M]_{2m} & \ddots&\cdots&\cdots & 0 & 0 \\
        \vdots & \ddots & \ddots&\cdots&\cdots & \vdots & \vdots\\
        0& 0 &\cdots&\cdots&\cdots& [M]_{2m} & [0]_{2m}\\
        0 & 0 & \cdots&\cdots&\cdots & 0  & [M]_{2m}\\
  \hline
        a_{1,1,1} & -a_{1,1,1} &\cdots  & -a_{1,m,1} & a_{1,1,2} &\cdots & -a_{1,m,p}\\
        \vdots&\vdots&\vdots&\vdots&\vdots&\vdots&\vdots\\
        a_{n,1,1} & -a_{n,1,1} &\cdots & -a_{n,m,1} & a_{n,1,2} &\cdots & -a_{n,m,p}\\
\end{bNiceArray}
\end{align*}
Note that this definition is in transpose form. The first $p$ rows of $W^{[1]}$ have $[M]_{2m}$ in a diagonal pattern at an indexes of $2m(k-1)+1$ through $2mk$ for each row $k$ where $k\in\{1, 2, \cdots, p\}$. The rest of the elements in the first $p$ rows are $0$. The remaining $n$ rows of $W^{[1]}$ consist of the coefficients for each equation in the subsystems with alternating signs as shown in Eqs. \ref{lem16 pos}-\ref{lem16 neg}.
\begin{equation*}
    B^{[1]} = -M +\begin{bmatrix}
        b_{1,1}&-b_{1,1} &\cdots&b_{m,1}&-b_{m,1}&\cdots&-b_{m,p}
    \end{bmatrix}^T
\end{equation*}
\begin{equation*}
    W^{[2]}=\begin{bmatrix}
        1&-1&0&0&\cdots&0&0\\
        0&0&1&-1&\ddots&0&0\\
        \vdots&\vdots&\ddots&\ddots&\ddots&\vdots&\vdots\\
        0&0&0&0&\cdots&1&-1\\
    \end{bmatrix}_{\times p}, B^{[2]}=0
\end{equation*}
\begin{align*}
    \text{where }& M\geq |f_k(x)|\text{ }\forall k \in \{1, 2, \cdots, p\}\text{ }\forall x\in Dom(f(x))\\
    W^{[1]T}&\in\mathbb{R}^{(p+n)\times 2mp},\text{ }B^{[1]}\in\mathbb{R}^{2mp},\text{ }W^{[2]}\in\mathbb{R}^{m\times 2mp}, \\
    [mat]_{\times p} &\text{ indicates repeating }[mat]\text{ horizontally p times.}
\end{align*}
Solving for $Y^{[1]}$ gives:
\begin{align*}
    Y^{[1]} &= \sigma^{[1]}(W^{[1]}X+B^{[1]})\\
    &=\sigma^{[1]}(\begin{bmatrix}
        M\mathbbm{1}_1(x)+\sum_{i=1}^n(a_{i,1,1}x_i)+b_{1,1}-M\\
        M\mathbbm{1}_1(x)+\sum_{i=1}^n(-a_{i,1,1}x_i)-b_{1,1}-M\\
        \vdots\\
        M\mathbbm{1}_1(x)+\sum_{i=1}^n(a_{i,m,1}x_i)+b_{m,1}-M\\
        M\mathbbm{1}_1(x)+\sum_{i=1}^n(-a_{i,m,1}x_i)-b_{m,1}-M\\
        M\mathbbm{1}_2(x)+\sum_{i=1}^n(a_{i,1,2}x_i)+b_{1,2}-M\\
        M\mathbbm{1}_2(x)+\sum_{i=1}^n(-a_{i,1,2}x_i)-b_{1,2}-M\\
        \vdots\\
        M\mathbbm{1}_p(x)+\sum_{i=1}^n(a_{i,m,p}x_i)+b_{m,p}-M\\
        M\mathbbm{1}_p(x)+\sum_{i=1}^n(-a_{i,m,p}x_i)-b_{m,p}-M\\
    \end{bmatrix})\\
    &=\sigma^{[1]}(\begin{bmatrix}
        M\mathbbm{1}_1(x)+u_{1,1}-M\\
        M\mathbbm{1}_1(x)-u_{1,1}-M\\
        \vdots\\
        M\mathbbm{1}_1(x)+u_{m,1}-M\\
        M\mathbbm{1}_1(x)-u_{m,1}-M\\
        M\mathbbm{1}_2(x)+u_{1,2}-M\\
        M\mathbbm{1}_2(x)-u_{1,2}-M\\
        \vdots\\
        M\mathbbm{1}_p(x)+u_{m,p}-M\\
        M\mathbbm{1}_p(x)-u_{m,p}-M\\
    \end{bmatrix})\\
    \end{align*}
    \begin{align*}
    &=\begin{bmatrix}
        \max(M\mathbbm{1}_1(x)+u_{1,1}-M,0)\\
        \max(M\mathbbm{1}_1(x)-u_{1,1}-M,0)\\
        \vdots\\
        \max(M\mathbbm{1}_1(x)+u_{m,1}-M,0)\\
        \max(M\mathbbm{1}_1(x)-u_{m,1}-M,0)\\
        \max(M\mathbbm{1}_2(x)+u_{1,2}-M,0)\\
        \max(M\mathbbm{1}_2(x)-u_{1,2}-M,0)\\
        \vdots\\
        \max(M\mathbbm{1}_p(x)+u_{m,p}-M,0)\\
        \max(M\mathbbm{1}_p(x)-u_{m,p}-M,0)\\
    \end{bmatrix}
\end{align*}
Solving for $Y^{[2]}$ gives:
\begin{align*}
    Y^{[2]} &= \sigma^{[2]}(W^{[2]}Y^{[1]}+B^{[2]})\\
    &=W^{[2]}Y^{[1]}\\
    &=\begin{bmatrix}
        1&-1&0&0&\cdots&0&0\\
        0&0&1&-1&\ddots&0&0\\
        \vdots&\vdots&\ddots&\ddots&\ddots&\vdots&\vdots\\
        0&0&0&0&\cdots&1&-1\\
    \end{bmatrix}_{\times p}\\
    &\times\begin{bmatrix}
        \max(M\mathbbm{1}_1(x)+u_{1,1}-M,0)\\
        \max(M\mathbbm{1}_1(x)-u_{1,1}-M,0)\\
        \vdots\\
        \max(M\mathbbm{1}_1(x)+u_{m,1}-M,0)\\
        \max(M\mathbbm{1}_1(x)-u_{m,1}-M,0)\\
        \max(M\mathbbm{1}_2(x)+u_{1,2}-M,0)\\
        \max(M\mathbbm{1}_2(x)-u_{1,2}-M,0)\\
        \vdots\\
        \max(M\mathbbm{1}_p(x)+u_{m,p}-M,0)\\
        \max(M\mathbbm{1}_p(x)-u_{m,p}-M,0)\\
    \end{bmatrix}\\
     \end{align*}
\vspace{-1.3cm}
\begin{equation*}
\ensuremath{
    =\begin{bmatrix}
        \sum_{k=1}^{p}(\max(M\mathbbm{1}_k(x)+u_{1,k}-M,0)-\max(M\mathbbm{1}_k(x)-u_{1,k}-M,0))\\
        \vdots\\
        \sum_{k=1}^{p}(\max(M\mathbbm{1}_k(x)+u_{m,k}-M,0)-\max(M\mathbbm{1}_k(x)-u_{m,k}-M,0))\\
    \end{bmatrix}
    }
\end{equation*}
Applying the assumption that $x$ is contained within only one subdomain, letting $x\in S_l$ implies $\mathbbm{1}_l=1$ while $\mathbbm{1}_k=0$ for all $k\in P= \{1, 2, \cdots, p\}$ and $k \neq l$. Recall that $M\geq|u|$ for all $x\in Dom(f(x))$. This gives:
\begin{align*}
   &Y^{[2]} =
   \begin{bmatrix}
        \max(u_{1,l},0)-\max(-u_{1,l},0)\\
        \vdots\\
        \max(u_{m,l},0)-\max(-u_{m,l},0)\\
    \end{bmatrix}\\
    &=\begin{bmatrix}
        \begin{cases}
        u_{1,l}, & u_{1,l}>0\\
        u_{1,l}, &u_{1,l}<0\\
        0, &u_{1l}=0\\
    \end{cases}\\
    \begin{cases}
        u_{2,l}, & u_{2,l}>0\\
        u_{2,l}, &u_{2,l}<0\\
        0, &u_{2,l}=0\\
    \end{cases}\\
    \vdots\\
    \begin{cases}
        u_{m,l}, & u_{m,l}>0\\
        u_{m,l}, &u_{m,l}<0\\
        0, &u_{m,l}=0\\
    \end{cases}\\
    \end{bmatrix}\\
    &=u_l
\end{align*}
Recall that if $x\in S_l$ then $f(x) = u_l$
\begin{equation}
    \therefore Y^{[2]} = f(x) : x\in Dom(f(x)), \sum_{k=1}^p\mathbbm{1}_k=1
\end{equation}
\end{pf*}

\subsection{Full Theorem}
\begin{thm}
\label{fullTH}
Given any continuous piecewise affine function, $f(x)=u$, with input $x$ of dimension $n$ and output $u$ of dimension $m$. Assume that $f(x)$ is defined on $p$ compact convex polytopes as subdomains, $S_k=Dom(f_k(x)), \, \forall k \in \{1, 2, \cdots, p\}$, that form a compact domain, $S=Dom(f(x))$. $f(x)$ can be expressed by a five-layer neural network without any training, given that no subdomains overlap and that $x\in S$.
\end{thm}

\begin{figure}[h]
    \centering
    \includegraphics[width=\linewidth]{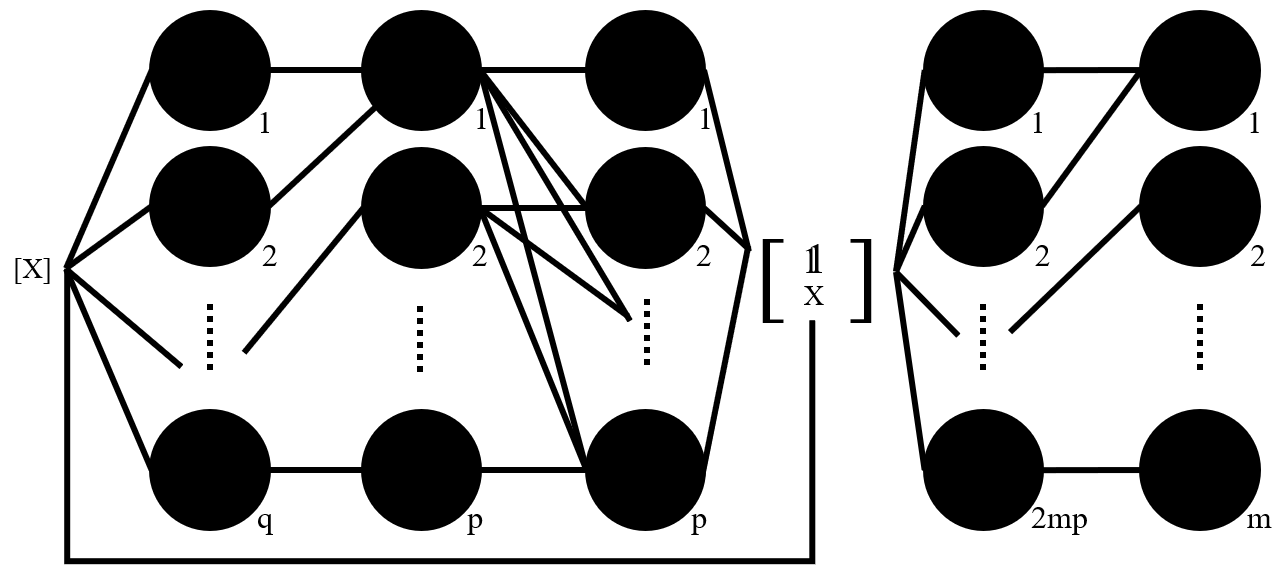}
    \caption{Full YANN architecture. }
    \label{fig:YANN}
\end{figure}

\begin{pf*}{Proof.} The half-space representations for each of the $p$ compact convex polytopes can be represented by a linear system of inequalities as written in Eq. \ref{eq:lem3 system s} where $q$ is the number of inequalities. The subfunctions of $f(x)=u$ can be represented by $p$ systems of equations as written in the two equivalent ways given in Eqs. \ref{lem16 pos}-\ref{lem16 neg}. Consider a five-layer neural network with input $X = [x_1, x_2, \cdots, x_n]^T$: layers one and two is the network presented in the proof of Lemma \ref{Lemma4}, layer three is the network presented in the proof of Lemma \ref{Lemma6}, and layers four and five is the network presented in the proof of Lemma \ref{Lemma 17}. It follows from Lemma \ref{Lemma4} that the output to layer two gives:

\begin{equation*}
    Y^{[2]} = \begin{bmatrix}
        \mathbbm{1}_1(x)\\
        \mathbbm{1}_2(x)\\
        \vdots\\
        \mathbbm{1}_p(x)
    \end{bmatrix}
\end{equation*}
Using $Y^{[2]}$ as the input to layer three, it follows from Lemma \ref{Lemma6} that the output gives:
\begin{equation*}
    Y^{[3]} = \begin{bmatrix}
        0\\
        \vdots\\
        \mathbbm{1}_k(x)=1\\
        \vdots\\
        0\\
    \end{bmatrix}
\end{equation*}
where $k$ is the first index in $\{1, 2, \cdots, p\}$ such that $x \in S_k$. Using $Y^{[3]}$ concatenated with $X$ as the input to layer four, i.e. $X^{[4]}=[Y^{[3]}; X]$, it follows from Lemma \ref{Lemma 17} that the output of the network gives:
\begin{equation*}
    Y^{[5]} = f(x) : x\in Dom(f(x))
\end{equation*}
Therefore the neural network architecture presented exactly represents the function $f(x)$.
\end{pf*}
\begin{rem}
    The concatenation to create the input for layer four is a skip or residual connection to use the original input of the network.
\end{rem}
\begin{rem}
    The assumption that $\sum_{k=1}^p\mathbbm{1}_k=1$ which was used in Lemma \ref{Lemma 17} is no longer needed since layer three of the network directly ensures this. 
\end{rem}
\begin{rem}
    The continuity of the affine function is crucial since polytopic subdomains may be connected through facets or points. If an input lies on a connected point or facet, more than one indicator function would evaluate to $1$ as more than one domains are simultaneously satisfied. Layer three of the network leads to the evaluation of whichever subfunction is indexed first in this scenario. The continuity implies that evaluating any subfunction for any of the subdomains on a connected point or facet gives the identical results, so it serves the same purpose whichever subfunction is evaluated.  
\end{rem}
\begin{rem}
    We take this full architecture to be the Y-wise Affine Neural Network (YANN) where layers one through three are the constraint checking section of the YANN and layers four through five are the function evaluation section of the YANN. A graphic of the YANN is provided in Fig. \ref{fig:YANN} where only non-zero weights are depicted as connections between nodes. 
\end{rem}
\begin{rem}\label{rem:24}
    The inclusion of the big $M$ constant to bound the ReLU outputs creates a mathematically exact representation. However, in practice, it is possible that an $M$ value is large enough to cause precision errors in computing. We have developed an alternative network, termed YANN-L, that avoids ReLU and $M$-bounding but it is much slower than the original YANN. The proof for the YANN-L is provided in Appendix \ref{YANN-L}.
\end{rem}
\section{Numerical case studies}
In this section, we demonstrate the computational benefits of YANNs on two examples for the control of dynamic systems. Example 2 considers a simple discrete-time, linear time invariant (LTI)  model. Example 3 considers a safety-critical reactor process. All case studies are performed on an Alienware m16 laptop with an Intel i9-13900HX CPU and an NVIDIA GeForce RTX 4090 Laptop GPU.
\begin{pf*}{Example 2.}
    This example is adapted from \citep{pistikopoulosMultiparametricOptimizationControl2020.ch10}. Consider the regulation MPC problem of the form given in Eq. \ref{eq:mpc} using the discrete-time double integrator presented in Eq. \ref{eq:LTI example}.
    \begin{equation}
        \label{eq:LTI example}
        A=\begin{bmatrix}
            1&1\\0&1
        \end{bmatrix} \quad
        B=\begin{bmatrix}
            0\\1
        \end{bmatrix} \quad
        C=\begin{bmatrix}
            1 &2
        \end{bmatrix} \quad
        D=0
    \end{equation}
\end{pf*}
With state path constraints $-10\leq x\leq10$, control input constraints $-1\leq u\leq 1$, output constraints $-25\leq y \leq 25$, $Q=I_2$, $R=0.01$, $N=10$. This MPC problem is reformulated into an mp-QP problem (Eq. \ref{eq:mp-QP mpc}) and solved to obtain the explicit control laws using the Python Parametric Optimization Toolbox (PPOPT) \cite{kenefake2022ppopt}. In this case, the optimal control input (Eq. \ref{eq:mp-QP soln}) is obtained as a piecewise affine function of the first measured system state over 115 subdomains (or critical regions). By using the approach presented in the proof of Theorem \ref{fullTH}, this piecewise affine solution is reformulated into a YANN. The evaluation speed up using the YANN is highlighted in Table \ref{tbl:example 2} where it evaluates an order of magnitude faster than traditional mp-MPC using PPOPT. The YANN shows a high degree of accuracy with a maximum error of 7.34E-6 over 1000 randomly generated initial states (Table \ref{tbl:example 2}). These results are performed on a CPU using FP32 precision where the inference time is measured for 1000 random states. The error can be reduced to 1E-14 using FP64 precision, but the packages used to optimize the network for inference speed do not support any precision higher than FP32. 
\begin{table}[h!]
\centering
\caption{Comparison of mp-MPC and YANN for 115 subdomains.}
\label{tbl:example 2}
\begin{tabular}{|*{6}{c|}}
  \hline
  \multicolumn{1}{|c|}{} & \multicolumn{1}{c|}{mp-MPC} & \multicolumn{1}{c|}{YANN} \\
  \hline
  1000 Inferences (s)&0.0223  &0.0077   \\
  \hline
  Avg. Time ($\mu$s)&22.3  &7.7   \\
  \hline
  Maximum Error&---  & 7.34E-6 \\
  \hline
\end{tabular}
\end{table}
\begin{pf*}{Example 3.}
    This example considers a continuous stirred tank reactor (CSTR) conceptualized from a real-world process safety incident \citep{braniffHierarchicalMultiparametricProgramming2024b,ALI2023100096}. The CSTR is described by the following dynamic equations, Eq. \ref{eq:cstr}, with the remaining modeling information and parameter definitions provided in Appendix \ref{CSTR Modeling}.
    \begin{subequations}\label{eq:cstr}
    \begin{equation}
        \frac{dC_{A}}{dt}=\frac{F_{A,in}-q_{out}}{V}C_A-k_{1}C_{A}C_{B}
    \end{equation}
    \begin{equation}
        \frac{dC_{B}}{dt}=\frac{F_{B,in}-q_{out}}{V}C_A-k_{1}C_{A}C_{B}
    \end{equation}
    \begin{equation}
        \frac{dC_{S}}{dt}=\frac{F_{C,in}-q_{out}}{V}C_A-k_{2}C_{S}
    \end{equation}
    \begin{equation}
         \frac{dT}{dt}=\frac{q_{out}}{V}(T_{in}-T)+\frac{\sum(-\Delta H_{k}r_{k})-\frac{UA_x}{V}(T-T_{c})}{\rho C_{P}}
    \end{equation}
\end{subequations}
Hydrogen, a highly flammable gas, is a byproduct in the two reactions taking place in this system which may cause safety concerns. Additionally, thermal runaway occurs if the reactor temperature reaches above $480$ K. Therefore, it is necessary to control the temperature below this threshold during operation in order to avoid thermal runaway which can lead to an explosion. Eq. \ref{eq:cstr} is linearized using the Jacobian method to give the following linear state-space model in Eq. \ref{eq:cstr-statespace} with states being concentrations, $C_A\text{, } C_B\text{, } C_S\text{,}$ and temperature, $T$. We have applied setpoint tracking mp-MPC to operate this safety-critical system in our prior work by manipulating the heat transfer coefficient, $U$, to control the temperature $T$ at different setpoints \citep{braniffHierarchicalMultiparametricProgramming2024b,ALI2023100096}. However, for simplicity, we consider only regulation in this example.
\begin{align*}
    &A = \begin{bmatrix}
         0.9506& 0& 0& 0\\
        -0.0484& 0.9943& 0& 0\\
        0& 0& 0.9909& 0\\
        0.6970& 0.0678& 0& 1.0030\\
    \end{bmatrix}
    B=\begin{bmatrix}
        0\\0\\0\\-0.0007
    \end{bmatrix}\\
\end{align*}
\vspace{-1cm}
\begin{align}
\label{eq:cstr-statespace}
    &C = \begin{bmatrix}
        0&0&0&1
    \end{bmatrix} \hspace{2.1cm} D=\begin{bmatrix}
        0
    \end{bmatrix}
\end{align}
This linear state-space model is used to formulate an MPC problem following Eq. \ref{eq:mpc} with $Q=I_4\times10^{4}$, $R=1\times10^{-6}$, $
        [\text{-}10\text{ -}10\text{ -}10\text{ -}20]\leq x_k^T \leq 
        [10\text{ } 10\text{ } 10\text{ } 20]$, $-25\leq y_k \leq 25$, and $-55 \leq u_k \leq 55$. The MPC problem is reformulated into an mp-QP problem following Eq. \ref{eq:mp-QP mpc} and solved via multi-parametric programming to obtain an explicit solution to the control problem as a function of the process states. In order to highlight the capability of YANNs across a varying amount of subdomains, we solve this problem for two cases: (i) a small operating and control horizon ($N=2$) which gives 9 subdomains, and (ii) for a longer horizon ($N=13$) which results in 2,615 subdomains.

The performance benefits of using YANNs are highlighted in Table \ref{tbl:CSTR speed}. A speed increase is noted over the traditional mp-MPC evaluation using PPOPT, regardless of the number of subdomains. The inference on CPU is quicker when the number of subdomains is small. This is because the network evaluates faster than the amount of time it takes to send information to the GPU. Once the CPU inference slows down enough the GPU offers an advantage. The blank cells in Tables \ref{tbl:CSTR speed}-\ref{tbl:CSTR error} indicate that no performance evaluation is performed. For example, in the case of 9 subdomains, it is not desired to evaluate the YANN on a GPU due to the afore-mentioned reason.

For the second case with significantly more subdomains, a decrease in accuracy is noted in the YANN. This arises from the use of the big $M$ parameter which causes computing precision loss as noted in Remark \ref{rem:24}. This issue is negligible for this example since 1E-2 is smaller than the control input by orders of magnitude ($-55 \leq u_k \leq 55$). If higher precision is desired, the YANN-L can be used for this purpose as shown in Table \ref{tbl:CSTR error}. However, YANN-L evaluates slower than YANN which presents a trade-off between accuracy and computational speed. 

\begin{table}[h]
\centering
\caption{mp-MPC and YANN performance comparison.}
\label{tbl:CSTR speed}
\begin{tabular}{|*{6}{c|}}
  \hline
  \multicolumn{1}{|c|}{Avg. Time ($\mu$s)} & \multicolumn{1}{c|}{mp-MPC} & \multicolumn{1}{c|}{CPU}  & \multicolumn{2}{c|}{GPU} \\
  \hline
    Subdomains &  & YANN  & YANN & YANN-L \\
  \hline
  9&14.2  &3.1   & ---& ---  \\
  \hline
  2,615&297.4  & --- &180.1 &630.0  \\
  \hline
\end{tabular}
\end{table}

\begin{table}[h]
\centering
\caption{YANN and YANN-L maximum error comparison.}
\label{tbl:CSTR error}
\begin{tabular}{|*{6}{c|}}
  \hline
  \multicolumn{1}{|c|}{Max Error}  & \multicolumn{2}{c|}{CPU}  & \multicolumn{2}{c|}{GPU} \\
  \hline
Subdomains    &  YANN & YANN-L  & YANN & YANN-L \\
  \hline
  9&$6.1E\text{-}6$  & --- & ---& ---  \\
  \hline
  2,615&$6.3E\text{-}3$  &$6.1E\text{-}6$  &$1.1E\text{-}2$ &$5.6E\text{-}6$   \\
  \hline
\end{tabular}
\end{table}

\end{pf*}
\section{Conclusions}

In this work we have introduced YANNs to exactly represent continuous piecewise affine functions of any input and output defined on any amount of domains without needing any training. We have applied this specialized architecture to efficiently represent the solutions to mp-MPC problems which inherently guarantees recursive feasibility and safety by being functionally equivalent to the explicit control policy. This is a first-of-its-kind NN-based controller since it requires no additional steps to develop or validate these control-theoretic properties. We have shown that YANNs evaluate faster than the traditional approaches for piecewise affine functions which enables faster control and the control for larger systems. We demonstrated examples to nonlinear functions or functions defined on non-linearly constrained domains. Ongoing work will look to leverage the YANNs interpretability in training algorithms. For example, it could be used in transfer learning to reduce network training time \citep{xiaoModelingPredictiveControl2023}. We will also investigate applications which use the nonlinear extensions to YANNs such as in the solutions to multi-parametric mixed-integer quadratic programs (mp-MIQPs) which can have quadratic constraints on subdomains \citep{narcisoMultiparametricMixedIntegerQuadratic2007}. 

\begin{ack}            
The authors gratefully acknowledge financial support from  NSF GRFP No. DGE-1102689, NSF RETRO Project CBET-2312457, and Department of Chemical and Biomedical Engineering at West Virginia University.
\end{ack}

\bibliographystyle{unsrt}        
\bibliography{bibliography}
\appendix
\section{Corollary to Lemma \ref{Lemma8}}
\begin{cor}
    Any affine equation with input $x\in \mathbb{R}^n$ can be exactly represented by a single neuron if only the positive or negative range of outputs are of interest. 
\end{cor}
\begin{pf*}{Proof.}
From the network presented in Lemma \ref{Lemma8}, the first node of layer one represents the positive range of outputs by itself. The second node of layer one solely represents the negative range of outputs.
\end{pf*}
\begin{rem}
    This Corollary can be useful in reducing network size and complexity in certain situations. If the input to an affine equation is bounded on a closed domain then by the boundedness theorem for continuous functions defined on closed domains the output is also bounded. In this way, it is possible to shift all outputs to be of the same sign by adding or subtracting some constant to the affine equation. This new representation can more efficiently evaluate the function and once a result is obtained the original solution can be resolved by shifting back by the same constant.  
\end{rem}
\section{Bounded function theorem}
\label{Bounded function th}
Let $f(x)=u$ be a continuous function defined on a set $S$. If $S$ is compact then $f(x)$ is bounded on $S$ and thus there exists an $M$ such that $M\geq|f(x)|, \forall x\in S$. 
\section{Proof for YANN-L}
\label{YANN-L}
For the proof of YANN-L we need to slightly modify Lemma \ref{Lemma 17}.
\begin{lem}
    Any continuous piecewise affine function, $f(x)=u$, with input $x\in \mathbb{R}^n$ and output $u\in \mathbb{R}^m$ that is defined on $p$ compact subdomains, $S_k=Dom(f_k(x)), \quad\forall k \in \{1, 2, \cdots, p\}$, that form a compact domain, $S=Dom(f(x))$, can be expressed by a three-layer neural network without any training given the solution to the indicator functions for each subdomain and assuming the input, $x$, is contained within one and only one subdomain.
\end{lem}
\begin{pf*}{Proof.} 
    Consider a three-layer neural network with the first layer comprising $mp$ nodes, the second layer being an elementwise matrix multiplication between the corresponding indicator function solutions and the outputs of layer one, and the third layer comprising $m$ nodes. It has input $X = [x_1, x_2, \cdots, x_n]^T$ and activation functions, $\sigma^{[1]} = (x)$, $\sigma^{[3]} = x$. The weights and biases of layer one are defined below using Eq. \ref{lem16 pos}, the weights of layer three are defined below, and the bias of layer three is zero.
    \begin{equation*}
        W^{[1]}=\begin{bmatrix}
            a_{1,1,1}&\cdots& a_{n,1,1}\\
            \vdots&\vdots&\vdots\\
            a_{1,m,1}&\cdots&a_{n,m,1}\\
            a_{1,1,2}&\cdots& a_{n,1,2}\\
            \vdots&\vdots&\vdots\\
            a_{1,m,p}&\cdots&a_{n,m,p}\\
        \end{bmatrix}
        B^{[1]}=\begin{bmatrix}
            b_{1,1}\\
            \vdots\\
            b_{m,1}\\
            b_{1,2}\\
            \vdots\\
            b_{m,p}
        \end{bmatrix}
    \end{equation*}
    \begin{equation*}
        W^{[3]}=\begin{bmatrix}
            1&&[0]_{m-1}\\
            0&1&[0]_{m-2}\\
            \vdots&\ddots&\ddots\\
            [0]_{m-1}&&[1]
        \end{bmatrix}_{\times p}\in \mathbb{R}^{m\times mp}
    \end{equation*}
    Solving for $Y^{[1]}$ gives:
    \begin{align*}
        Y^{[1]}=\begin{bmatrix}
             u_{1,1}\\
            \vdots\\
            u_{m,1}\\
            u_{1,2}\\
            \vdots\\
            u_{m,p}
        \end{bmatrix}
    \end{align*}
    Solving for $Y^{[2]}$ gives:
    \begin{align*}
        Y^{[2]}=\begin{bmatrix}
             u_{1,1}\\
            \vdots\\
            u_{m,1}\\
            u_{1,2}\\
            \vdots\\
            u_{m,p}
        \end{bmatrix}\circ
        \begin{bmatrix}
            \mathbbm{1}_{1}\\
            \vdots\\
            \mathbbm{1}_{1}\\
            \mathbbm{1}_{2}\\
            \vdots\\
            \mathbbm{1}_{p}
        \end{bmatrix}
        =\begin{bmatrix}
             \mathbbm{1}_{1}u_{1,1}\\
            \vdots\\
             \mathbbm{1}_{1}u_{m,1}\\
             \mathbbm{1}_{1}u_{1,2}\\
            \vdots\\
            \mathbbm{1}_{p}u_{m,p}
        \end{bmatrix}
    \end{align*}
    Solving for $Y^{[3]}$ gives:
    \begin{align*}
        Y^{[3]}
        =\begin{bmatrix}
             \sum_{l=1}^p\mathbbm{1}_{p}u_{1,p}\\
            \vdots\\
             \sum_{l=1}^p\mathbbm{1}_{p}u_{m,p}
        \end{bmatrix}
    \end{align*}
    Applying the assumption that $x$ is contained within only one subdomain which implies $\sum_{k=1}^p\mathbbm{1}_k=1$, and letting $x\in S_l$ which implies $\mathbbm{1}_l=1$ and $\mathbbm{1}_k=0$ for all other $k\in P= \{1, 2, \cdots, p\}$ such that $l \notin P$ gives:
    \begin{align*}
        Y^{[3]}
        =\begin{bmatrix}
             u_{1,l}\\
            \vdots\\
             u_{m,l}
        \end{bmatrix}
    \end{align*}
    Recall that since $x\in S_l$ then $f(x) = u_l$
\begin{equation}
    \therefore Y^{[3]} = f(x) : x\in Dom(f(x)), \sum_{k=1}^p\mathbbm{1}_k=1
\end{equation}
Following a similar set of steps to those outlined in the proof of Theorem \ref{fullTH}, the overall structure of the YANN-L can be easily realized. 
\end{pf*}
\section{Information for Example 3}
\noindent \underline{Reaction 1:} 
\begin{center}
Methylcyclopentadiene (A) + Sodium (B) \\ $\xrightarrow{Diglyme (S)}$ Sodium Methylcyclopentadiene (C)  + Hydrogen (D)
\end{center} 
\noindent \underline{Reaction 2:} \\
\begin{center} Diglyme (S) $\xrightarrow{Sodium (B)} $ Hydrogen (D) + Byproduct
\end{center}
\label{CSTR Modeling}
\begin{table}[ht]\label{tbl:CSTR_var}
  \centering
  \caption{List of modeling variables and parameters.}
  \begin{tabular}{|l|l|}
  \hline
  State variables & $C_A, C_B, C_S$: Concentrations\\& $T$: Temperature \\ \hline
  Manipulated variable & $U$: Heat transfer coefficient \\ \hline
  Control variable & $T$: Temperature \\ \hline
  Process parameters & $V$: Volume (4000 L)\\ &$\rho$: Mixture density (36 mol/L) \\ 
   & $C_p$: Specific heat (430.91 J/mol$\cdot$K) \\ 
  & $A_x$: Heat transfer area (5.3 m\textsuperscript{2})\\& $T_c$: Coolant temperature(373K) \\
  & $\Delta H_{k}$: (-45.6 kJ/mol, -320 kJ/mol)\\ 
  & $k_i = A_i\mbox{exp}(-\frac{E_i}{RT})$\\
  & $A_i$:($A_1 = 4\times10^{14}$, $A_2 = 1\times10^{84}$) \\
  & $E_i$:($E_1 = 1.28\times10^{5}$, $E_2 = 8\times10^{5}$ J/mol$\cdot$K) \\
  \hline
  \end{tabular} 
\end{table}                               
\end{document}